\documentclass[12pt]{article}
\usepackage{a4wide}

\newcommand{\DD}{{\cal D}}
\newcommand{\EE}{\hbox{\it I\kern-3pt E}}
\newcommand{\LL}{{\cal L}}
\newcommand{\OO}{{\cal O}}
\newcommand{\PP}{{\cal P}}
\newcommand{\TT}{{\cal T}}
\newcommand{\RR}{\hbox{\it I\kern-3pt R}}
\newcommand{\VV}{{\cal V}}
\newcommand{\pfrac}[2]{\left(\frac{#1}{#2}\right)}
\newcommand{\Ad}{\mathop{\rm Ad}\nolimits}
\newcommand{\dirvec}{\rlap{\kern-3pt\raise2pt\hbox{$^{^{\ \rightarrow}}$}}}
\newcommand{\revvec}{\rlap{\kern-3pt\raise2pt\hbox{$^{^{\ \leftarrow}}$}}}
\newcommand{\dubvec}{\rlap{\kern-3pt\raise2pt\hbox{$^{^{\ \leftrightarrow}}$}}}
\newcommand{\slD}{\kern1pt/\kern-8pt D}
\newcommand{\slF}{F\kern-8pt/\kern2pt}
\newcommand{\slG}{G\kern-8pt/\kern2pt}
\newcommand{\slk}{/\kern-6pt k}
\newcommand{\oone}{\hbox{\rm 1\kern-3pt l}}

\begin{document}

\begin{flushright}
MZ-TH/09-30\\
arXiv:0908.3761\\
June 2011
\end{flushright}
 
\begin{center}
\renewcommand{\thefootnote}{\fnsymbol{footnote}}
{\Large\bf ``Dynamical'' interactions and gauge invariance}

\vspace{24pt}
{\large R.~Saar$^1$, S.~Groote$^{1,2}$, 
  I.~Ots$^1$\footnote[2]{Deceased.} and H.~Liivat$^1$}\\[12pt]
$^1$ Loodus- ja Tehnoloogiateaduskond, F\"u\"usika Instituut,\\
  Tartu \"Ulikool, Riia~142, 51014 Tartu, Estonia\\[12pt]
$^2$ Institut f\"ur Physik der Johannes-Gutenberg-Universit\"at,\\
  Staudinger Weg 7, 55099 Mainz, Germany
\vspace{24pt}
\renewcommand{\thefootnote}{\arabic{footnote}}
\end{center}


\begin{abstract}
In order to avoid known, long-standing problems with higher-spin interactions,
the electromagnetic field is introduced ``dynamically'' by using a
nonsingular, Lorentz-type transformation, acting as adjoint representation on
the Poincar\'e algebra of the free theory. In doing so, Lorentz transformation
and local gauge transformation are placed on the same foundations, leading to
the phase transition as a consequence of the gauge transformation. The
procedure is exemplified in the case of plane waves for the Dirac-type
equation and the Rarita--Schwinger equation.
\end{abstract}
\begin{center}
{\em published as Physical Review D84 (2011) 065022}
\end{center}
\newpage

\section{Introduction}
Understanding the higher-spin interactions\footnote{In this paper states with
spin one and higher are considered as higher-spin states. This concept is not
universally accepted. For part of the investigators ``higher spin'' means
$s\geq 3/2$. The specialists in supergravity updated the convention of the
higher spin to be even $s\geq 5/2$~\cite{Vasiliev:2001ur}. Nevertheless, at
least in the Standard Model the troubles start already from the value $s=1$.
Therefore it seems that the convention $s\geq 1$ as the higher-spin region is
more justified than the other ones.} is a long-standing problem. However, in
spite of its 70 years history, the main goal -- the construction of a
consistent higher-spin theory, even for the electromagnetic interaction
which ought to be the simplest case -- has not been achieved yet. 

\vspace{7pt}
The theory of higher-spin interactions has never belonged to the ``mainstream''
theories. The field has been cultivated by groups of enthusiasts. On the other
hand, the theory of higher-spin interactions is needed for solving many
mainstream problems. It is related to the Standard Model (SM) in several ways.
By introducing the massive spin-one gauge bosons into the theory one also
introduces higher-spin problems into the Standard Model. Difficulties appear
for instance in scattering processes with the charged gauge bosons $W^\pm$ in
the initial or final state, or in constructing three-vertex gauge boson
self-interactions. A consistent higher-spin interaction theory is also needed
in chromodynamics. Quantum Chromodynamics (QCD) does not yet allow one to
describe low-energy hadronic processes in terms of underlying quark-gluon
dynamics. Because of this one has to use a more phenomenological approach in
terms of hadronic fields. However, one of the basic problems here is the
treatment of hadrons with higher spins~\cite{Pascalutsa:2003vz}.

\vspace{7pt}
Also for theories beyond the SM one needs a better understanding of ordinary
higher-spin field theory. String theory for instance is free of many
higher-spin problems and due to this it is believed that it can consistently
describe quantum gravity. A reason behind this consistent behaviour is that
string theories contain an infinite tower of all spin states. But at the same
time serious troubles exist in the physical interpretation of the string
theories. The existence of a consistent higher-spin interaction theory would
help in better understanding the physics behind the string theory. It is
believed that if a breakthrough in understanding the basic problems of the
ordinary higher-spin field theory happens, it might become a fashionable
topic~\cite{Sorokin:2004ie}.

\vspace{7pt}
The investigations of higher-spin fields started in the 1930s of the last
century with papers by Dirac~\cite{Dirac:1936tg}, Wigner~\cite{Wigner:1939cj},
Fierz and Pauli~\cite{Fierz:1939ix}, and were followed by the works of Rarita
and Schwinger~\cite{Rarita:1941mf}, Bargmann and Wigner~\cite{Bargmann:1948ck},
and others~\cite{Bhabha:1945zz,Wild:1947zz,HarishChandra:1947zz,Corson:1982wi,%
Pursey:1965zz,Tung:1967zz,Wightman:1977uc}. The difficulties in higher-spin
physics revealed themselves when one tried to couple higher-spin fields to an
electromagnetic field. In the 1960s concrete defects of the higher-spin
interaction theory were found. Johnson and Sudarshan~\cite{Johnson:1960vt} and
Schwinger~\cite{Schwinger:1963zz} demonstrated that in the case of minimal
electromagnetic coupling some of the anticommutation relations become
indefinite. It appeared that the defects were also present on the classical
level. Velo and Zwanziger~\cite{Velo:1969bt} and Shamaly and
Capri~\cite{Shamaly:1972zu} showed that in an external electromagnetic field
there appeared acausal (superluminal) modes of propagation.\footnote{Earnestly,
as shown by Cox~\cite{Cox:1989zz} the constraint analysis leading to these
acausal pathologies is incomplete. On the contrary, in the complete constraint
analysis a new tier of constraints occurs for the critical external field
values, reducing the pathology to the field-induced change of the degrees of
freedom. Because of these field-dependent constraints the analysis of acausal
models is very complicated.} Afterwards other defects -- bad high-energy
behaviour of the amplitudes, various algebraic problems {\it etc.} -- were
found. Since the sixties of the last century much work was done to solve the
problems, but no result which one can call a breakthrough was obtained in the
framework of ordinary field theory. In the case of higher-spin electromagnetic
interactions investigations of the last two decades have moved in two
directions. One part of the community develops the theory on the ground of
the minimal electromagnetic coupling, and the other part searches for a
consistent theory by using nonminimal couplings. 

\subsection{Problems of higher-spin interactions}
Difficulties in higher-spin physics are generic to all field theoretical
descriptions of relativistic higher-spin particles. They are related to the
fact that the covariant higher-spin field has more components than necessary
to describe the spin degrees of freedom of the physical particle. To get rid
of redundant degrees of freedom one must set up constraints between the field
components. If the interactions are introduced consistently with the free
field theory, the number of independent field components remains unchanged.
Otherwise the free theory constraints may be violated and unphysical degrees
of freedom become involved. The possibility to construct consistent
higher-spin theories with gauge invariant couplings was first pointed out by
Weinberg and Witten~\cite{Weinberg:1980kq}. However, the realization of this
scenario is beset with difficulties. Even though certain progress in
understanding of a higher-spin interaction theory has been
made~\cite{Sorokin:2004ie,progress}, up to now no general prescription for the
construction of a consistent higher-spin field theory for any spin has been
found.

\vspace{7pt}
In order to put constraints on the field components of the free theory it is
reasonable to use symmetry principles. The interacting theory then has to obey
similar symmetry requirements as the corresponding free theory or, even
better, preserves the symmetries of the free theory. Space-time properties of
a system under consideration are due to symmetries under the Poincar\'e group.
In fact, the very definition and characterization of distinct species of
elementary particles are provided by the set of inequivalent irreducible
projective unitary representations of the space-time symmetry group
$\PP_{1,3}$, the Poincar\'e group. According to conventional understanding of
a particle, its physical states of definite mass and spin, labeled by the
moment $p^\mu$ and the helicity $\lambda$, arise from the irreducible
representation of this symmetry group. The irreducible unitary representations
of the Poincar\'e group are characterized by the eigenvalues of the two
Casimir operators $P^2$ and $W^2$ of the Lie algebra $p_{1,3}$,
\begin{equation}\label{eq09}
P^2|m,s\rangle=m^2|m,s\rangle,\qquad
W^2|m,s\rangle=-m^2s(s+1)|m,s\rangle.
\end{equation}
The independent components of the Pauli--Lubanski pseudovector $W^\mu$ with
$[P_\mu,W_\nu]=0$ form the Lie algebra of the little group of fixed momentum
$p^\mu$. For every irreducible unitary representation of the little group one
can derive a corresponding irreducible induced representation of the
Poincar\'e group labeled by $(m,s)$, i.e.\ by the eigenvalues of the Casimir
operators in Eq.~(\ref{eq09}). Notice that the procedure of deriving induced
representations~\cite{Niederer:1974ps,Mackey:1952zz} corresponds very well to
the physical idea of first determining the internal degrees of freedom (the
helicity) of the system and then all its possible states of motion. In this
sense the natural identification of elementary particle systems is the direct
geometric transition from space-time to the system under consideration.

\vspace{7pt}
The identification of elementary particle systems and irreducible
representations of the Poincar\'e group finds its physical limitations in the
description of interacting systems and internal quantum numbers of composite
systems. Since gauge symmetry is a fundamental concept in Quantum
Electrodynamics, all physical quantities and dynamical equations of particles
have to be gauge invariant. However, if gauge invariance is realized by minimal
coupling, Poincar\'e invariance is violated at least for the theory of
higher-spin fields ($s\ge 1$) in its realization as first-order equations. The
deficits occur both on the classical level (acausality and algebraic
inconsistency) as well as on the quantum level (indefiniteness of antimutation
relations). A lot of work has been done to solve these problems, and a
consistent model for the spin-3/2 field in its realization as second order
equation (projector formalism) is proposed by Napsuciale {\it et al.} in
Ref.~\cite{Napsuciale:2006wr}.

\vspace{7pt}
In summarizing, one can conclude that if the problem is investigated by group
theoretical methods of space-time symmetries of interacting systems,
symmetries of interacting systems lead to the general covariance group in case
of a charged particle moving in an external electromagnetic field. As a
consequence, the group theoretical definition of an elementary particle can be
extended to the case where an external field is present. Even though the
Poincar\'e group is not a subgroup of the general covariance
group~\cite{Giovannini:1977wi,Schrader:1972zd}, this point of view is of help
to solve the problem.

\vspace{7pt}
In this paper we use a higher-spin electromagnetic interaction theory
developed by us earlier, based on the ``dynamical'' representation of the
Poincar\'e algebra as a dynamical principle which leads to a nonminimal
coupling. The representations are constructed from the generators of the free
Poincar\'e algebra and the external field in such a way that the new,
field-dependent generators obey the commutation relations of free Poincar\'e
algebra. Introducing the interactions in this way preserves the Poincar\'e
symmetry of the free theory and, hopefully, also the number of degrees of
freedom of the free theory. The dynamical theory has achieved success in
constructing causal spin-$3/2$ equations~\cite{Saar:1999ez} and for justifying
the value of gyromagnetic ratio $g=2$ for any spin~\cite{Ots:2001xn}.

\vspace{7pt}
The paper is organized as follows. In Sec.~2 we set up our conventions
related to the Poincar\'e group $\PP_{1,3}$. In Sec.~3 we explain how the
electromagnetic field $A$ can be introduced by using a nonsingular
transformation $\VV(A)$, specified by Lorentz-type gauge invariance.
Realizations are shown for the particular example of plane waves, leading to
local phase transformation via $e^{iq\lambda}$. In Sec.~4 we treat the
Rarita--Schwinger equation. Section~5 contains our conclusions and an outlook
on future work.

\section{The Poincar\'e group}
Relativistic field theories are based on the invariance under the Poincar\'e
group $\PP_{1,3}$ (known also as the inhomogeneous Lorentz group
${\cal IL}$~\cite{Wigner:1939cj,Bargmann:1954gh,Fronsdal:1959zz,Shaw:1964zz,%
Joos:1962qq,Niederer:1974ps,Ohnuki:1988ai,Tung:1985na,Kim:1986wv}) This group
is obtained by combining Lorentz transformations $\Lambda$ and space-time
translations $a_T$,
\begin{equation}
(a,\Lambda)\equiv a_T\Lambda:\EE_{1,3}\ni
  x^\mu\to{\Lambda^\mu}_\nu x^\nu+a^\mu.
\end{equation}
The group's composition law
$(a_1,\Lambda_1)(a_2,\Lambda_2)=(a_1+\Lambda_1a_2,\Lambda_1\Lambda_2)$
generates the semidirect structure of $\PP_{1,3}$,
\[\PP_{1,3}=\TT_{1,3}\odot\LL\]
where $\TT_{1,3}$ is the Abelian group of space-time translations (i.e.\ the
additive group $\RR^4$) and $\LL=\{\Lambda:\det\Lambda=+1,{\Lambda^0}_0\ge 1\}$
is the proper orthochronous Lorentz group\footnote{In order to simplify the
notation, in the following we refer to the proper orthochronous Lorentz group
and proper orthochronous Lorentz transformation as Lorentz group and Lorentz
transformation, respectively.} acting on the Minkowski space
$\EE_{1,3}$ with metric
\[\eta_{\mu\nu}=\mathop{\rm diag}(1,-1,-1,-1).\]
The condition of the metric to be invariant under Lorentz transformations
$\Lambda$ takes the form
\begin{equation}\label{eq02}
{\Lambda^\mu}_\rho\eta_{\mu\sigma}{\Lambda^\sigma}_\nu=\eta_{\rho\nu}.
\end{equation}
In order to set up the conventions used in this paper, in the following we
deal with the properties of representations of the Lorentz group in more
detail.

\subsection{Transformation of covariant functions}
Under the Lorentz transformation $\Lambda\in\LL$ the covariant functions $\psi$
transform according to a representation $\tau(\Lambda)$ of the Lorentz
group~\cite{Wigner:1939cj,Bargmann:1948ck,Corson:1982wi,Pursey:1965zz,%
Tung:1967zz,Wightman:1977uc,Bargmann:1954gh,Fronsdal:1959zz,Shaw:1964zz,%
Joos:1962qq,Niederer:1974ps} where the diagram
\begin{center}\begin{tabular}{rlcr}
$\psi:$&$x\in\EE_{1,3}$&$\longrightarrow$&$\psi(x)$\\
$\llap{$\tau(\Lambda)$}\downarrow$&$\downarrow\rlap{$\Lambda$}$&&
  $\downarrow\rlap{$T(\Lambda)$}\quad$\\
$\tau(\Lambda)\psi:$&$\Lambda x$&$\longrightarrow$&$T(\Lambda)\psi(x)$\\
\end{tabular}\end{center}
is commutative, i.e.\
\begin{equation}\label{eq01}
\psi^\Lambda(\Lambda x)\equiv(\tau(\Lambda)\psi)(\Lambda x)
  =T(\Lambda)\psi(x).
\end{equation}
The map $T:\Lambda\to T(\Lambda)$ is a finite-dimensional representation of
$\LL$. If we parametrize the element $\Lambda\in\LL$ by
$\Lambda(\omega)=\exp\Big(-\frac12\omega_{\mu\nu}e^{\mu\nu})$ where the
Lorentz generators are given by
\[
{(e_{\mu\nu})^\rho}_\sigma=-{\eta_\mu}^\rho\eta_{\nu\sigma}+\eta_{\mu\sigma}
{\eta_\nu}^\rho
\]
and $\omega^{\mu\nu}=-\omega^{\nu\mu}$ are six independent parameters, the
parametrization of $T$ reads
\[
T(\Lambda(\omega))=\exp\Big(-\frac i2\omega_{\mu\nu}s^{\mu\nu}).
\]
The Lorentz group $\LL$ is noncompact. As a consequence, all unitary
representations are infinite dimensional. In order to avoid this, we introduce
the concept of $H$-unitarity (see e.g.\ Ref.~\cite{Niederer:1974ps} and
references therein). A finite representation $T$ is called $H$-unitary if
there exists a nonsingular hermitian matrix $H=H^\dagger$ so that
\begin{equation}\label{eq03}
T^\dagger(\Lambda)H=HT^{-1}(\Lambda)\quad\Leftrightarrow\quad
s_{\mu\nu}^\dagger H=Hs_{\mu\nu}.
\end{equation}
Notice that a $H$-unitary metric is always indefinite, so that the inner
product $\langle\ ,\ \rangle$ generated by $H$ is sesquilinear sharing the
Hermiticity condition
$\langle\psi,\varphi\rangle=\langle\varphi,\psi\rangle^*$. The most famous
case of $H$-unitarity is given in the Dirac theory of spin-1/2 particles
where $H=\gamma^0$.

\subsection{Transformation of operators}
The transformation $\tau(\Lambda)$ in Eq.~(\ref{eq01}) is a covariant
transformation for the operator $\OO$~\cite{Giovannini:1977wi,Janner:1970zz}
acting on the $\psi$-space of covariant functions\footnote{We have to impose
the action on covariant functions because in the case of higher spins the
relations between operators we obtain are valid only as weak conditions.} if
the diagramm
\begin{center}\begin{tabular}{rlcr}
$\OO\psi:$&$x$&$\longrightarrow$&$(\OO\psi)(x)$\\
$\llap{$\tau(\Lambda)$}\downarrow$&$\downarrow\rlap{$\Lambda$}$&
  &$\downarrow\ T(\Lambda)$\\
$\tau(\Lambda)(\OO\psi):$&$\Lambda x$&$\rightarrow$&$T(\Lambda)(\OO\psi)(x)$\\
\end{tabular}\end{center}
is commutative, i.e.\
\begin{equation}\label{eq05}
(\tau(\Lambda)\OO\tau^{-1}(\Lambda))(\Lambda x)(\tau(\Lambda)\psi)(\Lambda x)
  =T(\Lambda)\OO(x)\psi(x).
\end{equation}
Using Eq.~(\ref{eq01}) we obtain
\[(\tau(\Lambda)\OO\tau^{-1}(\Lambda))(\Lambda x)T(\Lambda)\psi(x)
  =T(\Lambda)\OO(x)\psi(x).\]
Notice that the covariance of the transformation embodies only the property of
equivalence of reference systems. The covariant operator $\OO$ is invariant
under the transformation~(\ref{eq01}) if in addition
$\OO\tau(\Lambda)=\tau(\Lambda)\OO$. As a consequence we obtain the
commutative diagram
\begin{center}\begin{tabular}{rlcr}
$\OO\psi:$&$x$&$\longrightarrow$&$(\OO\psi)(x)$\\
$\llap{$\tau(\Lambda)$}\downarrow$&$\downarrow\rlap{$\Lambda$}$&
  &$\downarrow\ T(\Lambda)$\\
$\OO(\tau(\Lambda)\psi):$&$\Lambda x$&$\longrightarrow$
  &$T(\Lambda)(\OO\psi)(x)$\\
\end{tabular}\end{center}
or
\begin{equation}\label{kom0}
\OO(\Lambda x)T(\Lambda)\psi(x)=T(\Lambda)\OO(x)\psi(x)
\end{equation}
which means
\begin{equation}\label{eq06}
\OO(\Lambda x)T(\Lambda)=T(\Lambda)\OO(x)
\end{equation}
on the $\psi$-space. The invariance is a symmetry of the physical system and
implies the conservation of currents. In particular, the symmetry
transformations leave the equations of motion form-invariant.

\subsection{The Lie algebra}
While the Lorentz transformation $T(\Lambda)$ changes the wave function $\psi$
itself as well as the argument of this function (cf.\ Eq.~(\ref{eq01})), the
proper Lorentz transformation $\tau(\Lambda)$ causes a change of the wave
function only. On the ground of infinitesimal transformations, this change is
performed by a substantial variation. Starting from an arbitrary
infinitesimal coordinate transformation
$\Lambda(\delta\omega):x^\mu\to x^\mu+\delta\omega^{\mu\nu}x_\nu$, the
substantial variation is given by Ref.~\cite{Corson:1982wi}
\[
\delta_0\psi(x)\equiv\psi^\Lambda(x)-\psi(x)
  =-\frac{i}2\delta\omega^{\rho\sigma}M_{\rho\sigma}\psi(x)
\]
where $M_{\rho\sigma}=\ell_{\rho\sigma}+s_{\rho\sigma}$, and
$\ell_{\rho\sigma}=i(x_\rho\partial_\sigma-x_\sigma\partial_\rho)$. The
corresponding finite proper Lorentz transformation can be written as
\[\tau(\Lambda(\omega))=\exp\left(-\frac{i}2\omega_{\mu\nu}M^{\mu\nu}\right),\]
and the multiplicative structure of the group generates the adjoint action
\begin{equation}\label{adjact}
\Ad_{\tau(\Lambda)}:M_{\mu\nu}\to\tau^{-1}(\Lambda)M_{\mu\nu}
  \tau(\Lambda)={\Lambda_\mu}^\rho{\Lambda_\nu}^\sigma M_{\rho\sigma}.
\end{equation}
Because of Eq.~(\ref{eq03}) the generators $s_{\rho\sigma}$ fulfill
$s_{\rho\sigma}^\dagger H=Hs_{\rho\sigma}$. They depend on the spin of the
field but not on the coordinates $x_\mu$. Therefore, we have
$[\ell_{\mu\nu},s_{\rho\sigma}]=0$. If a generic element of the translation
group is written as
\[\exp(+ia_\mu P^\mu),\]
the commutator relations of the Lie algebra are given by
\begin{eqnarray}\label{eq08}
[M_{\mu\nu},M_{\rho\sigma}]&=&i(\eta_{\mu\sigma}M_{\nu\rho}
  +\eta_{\nu\rho}M_{\mu\sigma}-\eta_{\mu\rho}M_{\nu\sigma}
  -\eta_{\nu\sigma}M_{\mu\rho}),\nonumber\\[7pt]
[M_{\mu\nu},P_\rho]&=&i(\eta_{\nu\rho}P_\mu-\eta_{\mu\rho}P_\nu),
  \nonumber\\[7pt]
[P_\mu,P_\nu]&=&0.
\end{eqnarray}
The Casimir operators of the algebra are $P^2=P_\mu P^\mu$ and
$W^2=W_\mu W^\mu$ where
\[W^\mu=+\frac12\epsilon^{\mu\nu\rho\sigma}M_{\nu\rho}P_\sigma\]
is the Pauli-Lubanski pseudovector, $[P_\mu,W_\nu]=0$. In coordinate
representation we have $P_\mu=i\partial_\mu$, and the finite Poincar\'e
transformation has the form
\begin{equation}
\tau(a,\Lambda):\psi(x)\to\left(\tau(a,\Lambda)\psi\right)(x)
  =T(\Lambda)\psi\left(\Lambda^{-1}(x-a)\right).
\end{equation}
This relation constitutes the Lorentz--Poincar\'e
connection~\cite{Tung:1985na}. While the representation $T$ generally
generates a reducible representation of $\PP_{1,3}$, the spectra of the
Casimir operators $P^2$ and $W^2$ determine the mass and spin content of the
system.

\subsection{Linear wave equation}
As a rule the Lorentz--Poincar\'e connection is realized by the relativistic
wave equations. If the relativistic wave equation transforms as a
finite-dimensional representation of the Lorentz group by Eq.~(\ref{eq01}), it
contains spins exceeding the desired physical spins. In order that the
solutions of the field equation correspond to a particle with a definite spin,
the equation must act like a projection operator to pick out the desired spin
components, i.e.\ to select the corresponding irreducible representation of
the Poincar\'e group.

\vspace{7pt}
The wave equation we consider has the form
\begin{equation}\label{eq10}
\DD(\partial)\psi(x)\equiv(i\gamma^\mu\partial_\mu-\rho)\psi(x)=0
\end{equation}
where $\psi$ is an $N$-component function, $\gamma^\mu$ ($\mu=0,1,2,3$), and
$\rho$ are $N\times N$ matrices independent of $x$. Following Bhabha's
conception~\cite{Bhabha:1945zz}, it is ``\dots\ logical to assume that the
fundamental equations of the elementary particles must be first-order
equations of the form~(\ref{eq10}) and that all properties of the particles
must be derivable from these without the use of any further subsidiary
conditions.''\footnote{In order to avoid confusion, we have to emphazise that
in citing Bhabha~\cite{Bhabha:1945zz} we do not imply Eq.~(\ref{eq10}) to be
the Bhabha equation. Instead, the equation is a generalization of the Dirac
equation, obeying the Lorentz conditions explained in the following.}
Therefore, different from Napsuciale's approach~\cite{Napsuciale:2006wr} we
start from first order differential equations. 

\vspace{7pt}
The principle of relativity states that a change of the reference frame cannot
have implications for the motion of the system. This means that
Eq.~(\ref{eq10}) is invariant under Lorentz transformations. Equivalently, the
Lorentz symmetry of the system means the covariance and form invariance of
Eq.~(\ref{eq10}) under the transformation in Eq.~(\ref{eq01}), i.e.\ the
transformed wave equation is equivalent to the old one. Therefore, we require
that every solution $\psi^\Lambda(\Lambda x)$ of the transformed equation
$\DD^\Lambda(\Lambda\partial)\psi^\Lambda(\Lambda x)=0$ can be obtained as
Lorentz transformation of the solution $\psi(x)$ of Eq.~(\ref{eq10}) in the
original system and that the solutions in the original and transformed systems
are in one-to-one correspondence. The explicit form of the covariance follows
from Eq.~(\ref{eq05}),
\begin{equation}\label{eq11}
\DD^\Lambda(\Lambda\partial)\psi^\Lambda(\Lambda x)
  =(\tau(\Lambda)\DD\tau^{-1}(\Lambda))(\Lambda\partial)
  (\tau(\Lambda)\psi)(\Lambda x)=T(\Lambda)\DD(\partial)\psi(x)=0,
\end{equation}
leading to the explicit Lorentz transformations
\[\gamma^{\prime\mu}={\Lambda^\mu}_\rho T(\Lambda)\gamma^\rho T^{-1}(\Lambda),
  \qquad\rho'=T(\Lambda)\rho T^{-1}(\Lambda).\]
The Lorentz invariance is given by the substitution
\[\DD(\partial)\psi(x)=0\
\buildrel{{\rm Eq.\,(\ref{eq01})}}\over\longrightarrow\
\DD(\partial)\psi^\Lambda(x)=0\]
or
\[T^{-1}(\Lambda)\gamma^\mu T(\Lambda)={\Lambda^\mu}_\rho\gamma^\rho,\qquad
T^{-1}(\Lambda)\rho T(\Lambda)=\rho.\]
The difference of the original and transformed wave equation is given by the
wave equation where the wave function $\psi$ is replaced by the substantial
variation $\delta_0\psi$, $\DD(\partial)\delta_0\psi(x)=0$. As a consequence
we obtain $[\DD,M^{\rho\sigma}]=0$ or
\begin{equation}\label{eq13}
[\gamma^\mu,s^{\rho\sigma}]=i(\eta^{\mu\rho}\gamma^\sigma
  -\eta^{\mu\sigma}\gamma^\rho),\qquad[\rho,s^{\rho\sigma}]=0.
\end{equation}
An excellent discussion of such matrices $\gamma$ can be found in
Refs.~\cite{Fierz:1939zz,Bhabha:1945zz,Wild:1947zz,Corson:1982wi,Gelfand:1963,%
Naimark:1964}. The Hermiticity of the representation $T$ in Eq.~(\ref{eq03})
implies the Hermiticity of Eq.~(\ref{eq10}). Including a still unspecified
Hermitian matrix $H$ the Hermiticity condition reads
$\DD(\partial)^\dagger H\buildrel!\over=(\DD(\partial)H)^\dagger
  =-H\DD(-\partial)$ or
\begin{equation}\label{eq14}
\gamma^{\mu\dagger}H=H\gamma^\mu,\qquad\rho H=H\rho.
\end{equation}
Writing $\bar\psi=\psi^\dagger H$, one obains the adjoint equation
\begin{equation}\label{eq15}
\bar\psi\DD(-\revvec\partial)=\bar\psi(-i\gamma^\mu\revvec{\partial_\mu}-\rho)
  =-(H\DD(\partial)\psi)^\dagger=0.
\end{equation}

\section{Introduction of the external field}
Because of the arguments given earlier, it may be reasonable to introduce the
external field directly into the Poincar\'e algebra. To do so one has to
transform the generators of the Poincar\'e group to be dependent on the
external field in such a way that the new, field-dependent generators obey the
commutation relations~(\ref{eq08}). As proposed by
Chakrabarti~\cite{Chakrabarti:1968zz} and Beers and Nickle~\cite{Beers:1972xt},
the simplest way to build such a field-dependent algebra is to introduce the
external field by a nonsingular transformation
\begin{equation}\label{defVV}
\Ad_{\VV(A)}:p_{1,3}\to p_{1,3}^d(A)=\VV(A)p_{1,3}\VV^{-1}(A)
  =p_{1,3}+[\VV(A),p_{1,3}]\VV^{-1}(A).
\end{equation}
More explicitely, the transformed operators
\begin{eqnarray}\label{eq2.1}
\Pi_\mu(A)&=&P^\mu+[\VV(A),P_\mu]\VV^{-1}(A),\nonumber\\[7pt]
\xi_\mu(A)&=&x^\mu+[\VV(A),x_\mu]\VV^{-1}(A),\nonumber\\[7pt]
\sigma_{\mu\nu}(A)&=&s_{\mu\nu}+[\VV(A),s_{\mu\nu}]\VV^{-1}(A),\nonumber\\[7pt]
\mu_{\mu\nu}(A)&=&\xi_\mu(A)\Pi_\nu(A)-\xi_\nu(A)\Pi_\mu(A)+\sigma_{\mu\nu}(A)
\end{eqnarray}
must satisfy the commutation relations of the Poincar\'e algebra.
In the case of a particular external electromagnetic field $A$, the external
field can be introduced by using an evolution operator $\VV(A)$, called the
dynamical representation~\cite{Saar:1999ez,Ots:2001xn}. By analogy with the
free particle case one can realize this representation on the solution space of
relativistically invariant equations. Expressing the operators explicitly in
terms of free-field operators, one obtains the dynamical interaction. Applying
for instance the operator $\VV(A)$ to Eq.~(\ref{eq10}) one obtains
\begin{equation}\label{eq3.1}
\VV(A):\DD(\partial)\psi(x)=0\quad\rightarrow\quad\DD^d(\partial,A)\Psi(x,A)=0
\end{equation}
where $\DD^d(\partial,A)=\VV(A)\DD(\partial)\VV^{-1}(A)$ and
\begin{equation}\label{eq3.2}
\Psi(x,A)=\VV(A)\psi(x)
\end{equation}
(here and in the following we skip the argument $x$ for $\Psi$ and the
argument $\partial$ for $\DD^d$). Having introduced the external gauge field
$A$, we introduce gauge covariance on the same foundation as Lorentz covariance
in Eq.~(\ref{eq01}), i.e.\ by claiming that the diagram
\begin{center}\begin{tabular}{rlcr}
$\Psi:$&$A$&$\longrightarrow$&$\Psi(A)$\\
$\llap{$g(\lambda)$}\downarrow\ $&$\ \downarrow\rlap{$\lambda$}$
  &&$\downarrow\rlap{$G(\lambda)$}$\\
$\Psi^\lambda:$&$A^\lambda=A+\partial\lambda$&$\longrightarrow$
  &$G(\lambda)\Psi(A)$\\
\end{tabular}\end{center}
is commutative,
\begin{equation}\label{kom1}
\Psi^\lambda(A+\partial\lambda)=G(\lambda)\Psi(A).
\end{equation}
According to Eq.~(\ref{kom0}), the dynamical interaction $\DD^d$ is gauge
invariant under the gauge transformation
$A\to A^\lambda\equiv A+\partial\lambda$ if the diagram
\begin{center}\begin{tabular}{rlcr}
$\DD^d\Psi:$&$A$&$\longrightarrow$&$\DD^d(A)\Psi(A)$\\
$\downarrow\quad$&
$\ \downarrow\rlap{$\lambda$}$&&$\downarrow\rlap{$G(\lambda)$}\qquad$\\
$\DD^d\Psi^\lambda:$&$A+\partial\lambda$&$\longrightarrow$
  &$G(\lambda)\DD^d(A)\Psi(A)$\\
\end{tabular}\end{center}
is commutative, i.e.\
\begin{equation}\label{kom2}
\DD^d(A+\partial\lambda)\Psi^\lambda(A+\partial\lambda)
  =G(\lambda)\DD^d(A)\Psi(A).
\end{equation}
Together with Eq.~(\ref{kom1}) we obtain
$\DD^d(A+\partial\lambda)G(\lambda)\Psi(A)=G(\lambda)\DD^d(A)\Psi(A)$ or
\begin{equation}\label{kom2p}
\DD^d(A+\partial\lambda)G(\lambda)=G(\lambda)\DD^d(A)
\end{equation}
on the $\psi$ space. Note that up to now we have not specified the explicit
shape of the finite-dimensional representation $G:\lambda\to G(\lambda)$ of
the gauge group.

\subsection{Specifying $\VV(A)$ by gauge invariance}
At this point we specify $\VV(A)$ by two claims~\cite{Saar:2010pc}. Because of
gauge symmetry as a fundamental principle the dynamical transformation $\VV(A)$
has to be compatible with the gauge transformation. Therefore, we first claim
the gauge invariance in Eq.~(\ref{kom2p}) not only for the operator $\DD^d$
but for the whole dynamical Poincar\'e algebra $p_{1,3}^d(A)$,
\begin{equation}
p_{1,3}^d(A+\partial\lambda)G(\lambda)=G(\lambda)p_{1,3}^d(A).
\end{equation}
By using Eq.~(\ref{defVV}) and multiplying by $G(\lambda)^{-1}$ from the right
we obtain
\begin{equation}
\VV(A+\partial\lambda)p_{1,3}\VV^{-1}(A+\partial\lambda)
  =G(\lambda)\VV(A)p_{1,3}(G(\lambda)\VV(A))^{-1}.
\end{equation}
This means that the first claim is fulfilled if
$\VV(A+\partial\lambda)=G(\lambda)\VV(A)$. On the other hand, with
Eqs.~(\ref{eq3.2}) and~(\ref{kom1}) we obtain
\begin{equation}
\VV^\lambda(A+\partial\lambda)\psi(x)=G(\lambda)\VV(A)\psi(x)
\end{equation}
and, therefore, $\VV^\lambda=\VV$ on the $\psi$-space. To summarize, by the
first claim the gauge symmetry determines the gauge properties of $\VV(A)$
and, therefore, the gauge properties of the interacting field $\Psi(A)$,
\begin{equation}\label{Psitrans}
\Psi(A)\to\Psi(A+\partial\lambda)=G(\lambda)\Psi(A).
\end{equation}

\vspace{7pt}
The second claim is that the dynamical transformation operator $\VV(A)$ should
be of Lorentz type, i.e.\ for the generators $s_{\mu\nu}$ of the Poincar\'e
algebra $p_{1,3}$ one has 
\begin{equation}\label{eq2.2}
\VV(A)s^{\mu\nu}\VV^{-1}(A)={V^\mu}_\rho(A){V^\nu}_\sigma(A)s^{\rho\sigma}
\end{equation}
which is a local extension of Eq.~(\ref{adjact}).
$V_{\mu\nu}(A)=V_{\mu\nu}(x,A)$ is the local Lorentz transformation generated
by the external field $A$ and obeying
\begin{equation}\label{eq2.3}
V_{\mu\rho}(A){V^\mu}_\sigma(A)
  =V_{\rho\mu}(A){V_\sigma}^\mu(A)=\eta_{\rho\sigma}.
\end{equation}
If such a local Lorentz transformation exists, the problem is solved.

\subsection{Solution for plane waves}
It is not easy to construct the Lorentz transformation $V_{\mu\nu}(A)$ in
general. In a sequel of this paper we deal with this problem in detail. In
order to show that such a solution exists, in the following we give an example
for an explicit realization of the local Lorentz transformation
$V_{\mu\nu}(A)$. As first shown by Taub~\cite{Taub:1969zz}, in the case of a
plane-wave field we obtain
\begin{equation}\label{eq2.4}
V_{\mu\nu}(A)=\eta_{\mu\nu}-\frac{q}{k_P}(k_\mu A_\nu-k_\nu A_\mu)
  -\frac{q^2}{2k_P^2}A^2k_\mu k_\nu
\end{equation}
where $q$ is the electric charge of the particle.\footnote{Note that the
plane-wave solution of the Dirac equation was found more than 70 years ago by
Volkov~\cite{Wolkow:1935} and extended later on to a field of two beams of
electromagnetic radiation~\cite{SenGupta:1967,Pardy:2005kv}. However, these
approaches did not make use of the nonsingular transformation $\VV(A)$.} The
plane wave is characterized by its lightlike propagation vector $k_\mu$,
$k^2=0$, and its polarization vector $a^\mu$ such that
\begin{equation}\label{eq2.5}
a^2=-1,\qquad ka=0.
\end{equation}
The operator $k_P\equiv k_\mu P^\mu$ commutes with any other operator and has
a special role in the theory. For particles with nonzero mass one has
$k_\mu P^\mu\ne 0$. Therefore, for the plane wave the operator $1/k_P$ is
well defined for the plane-wave solution $\psi_P$ of the Klein--Gordon
equation. In all other cases, $1/k_P$ is assumed to exist (for a further
discussion see the Appendix).

\vspace{7pt}
We write $A_\mu(\xi)=a_\mu f(\xi)$, where the variable $\xi=k_\mu x^\mu$ can
be used in place of the proper time. From Eq.~(\ref{eq2.5}) one obtains the
conditions
\begin{equation}\label{eq2.6}
\partial_\mu A^\mu=k_\mu\frac{dA^\mu(\xi)}{d\xi}=k_\mu A^{\mu\prime}(\xi)=0,
  \qquad k_\mu A^\mu=0
\end{equation}
where we used $A'_\mu(\xi)=dA_\mu(\xi)/d\xi$, while the field
\[F_{\mu\nu}=\partial_\mu A_\nu-\partial_\nu A_\mu
  =k_\mu A'_\nu(\xi)-k_\nu A'_\mu(\xi)=F_{\mu\nu}(\xi)\]
satisfies
\begin{eqnarray}\label{eq2.7}
\partial_\mu F^{\mu\nu}&=&k_\mu F^{\mu\nu}(\xi)\ =\ 0,\nonumber\\
F_{\mu\nu}{F^\nu}_\rho&=&-k_\mu k_\rho\left(A'(\xi)\right)^2.
\end{eqnarray}
It turns out that Eq.~(\ref{eq2.4}) can be written as
\begin{equation}
V_{\mu\nu}(A)=\exp\left(-\frac{q}{k_P}G\right)_{\mu\nu}
\end{equation}
where $G_{\mu\nu}=k_\mu A_\nu-k_\nu A_\mu$. Note that the exponential series
truncates after the second order term. In addition one obtains
\begin{eqnarray}\label{eq2.8}
V_{\mu\nu}(A)&=&V_{\nu\mu}(A)+\frac{2q}{k_P}G_{\nu\mu}\nonumber\\
V_{\mu\nu}(A)k^\nu&=&V_{\nu\mu}(A)k^\nu\ =\ k_\mu,\nonumber\\
\ [P_\mu,V_{\rho\sigma}(A)]&=&-i\frac{q}{k_P}k_\mu F_{\rho\sigma}
  -i\frac{q^2}{k_P^2}(AA')k_\mu k_\rho k_\sigma.
\end{eqnarray}
From the second equation in~(\ref{eq2.8}) one concludes that $V_{\mu\nu}(A)$
is an element of the (local) little group $\LL g(\xi)$ of the propagation
vector $k_\mu$. It is easy and interesting to see that $V_{\mu\nu}(A)$
generates a gauge transformation on $A_\mu$,
\begin{equation}\label{eq2.9}
V_{\mu\nu}(A)A^\nu=A_\mu+\partial_\mu\lambda_V(\xi),\qquad
\lambda_V(\xi)=-\frac{q}{k_P}\int_{\xi_0}^\xi d\xi'A^2(\xi'),
\end{equation}
and that the field $F_{\mu\nu}$ is invariant under this gauge transformation,
\begin{equation}\label{eq2.9i}
{V^\mu}_\rho(A){V^\nu}_\sigma(A)F^{\rho\sigma}=F^{\mu\nu}.
\end{equation}
Therefore, the local Lorentz transformation $V_{\mu\nu}(A)$ is a symmetry.
Notice that the local Lorentz transformation~(\ref{eq2.4}) has been rederived
many times~\cite{Brown:1964zz,Kupersztych:1978dq,Brown:1984hy} and widely
exploited often in the context of its physical implications. In particular, at
the classical level the solutions of the Lorentz form equation can be
expressed in terms of these local transformations~(\ref{eq2.4}). Therefore, in
the plane-wave case $V_{\mu\nu}(A)$ plays the role of an evolution operator.

\vspace{7pt}
The realization of $\VV(A)$ can be achieved by the nonsingular transformation
$\VV(A)=\VV_0(A)\VV_s(A)$ where
\begin{eqnarray}\label{eq2.10}
\VV_0(A)&=&\exp\Bigg\{-i\int\frac{d\xi}{2k_P}(2q(AP)-q^2A^2)\Bigg\},\nonumber\\
\VV_s(A)&=&\exp\Bigg\{-\frac{iq}{2k_P}G_{\mu\nu}s^{\mu\nu}\Bigg\}.
\end{eqnarray}
It has to be mentioned that the evolution operator $\VV(A)$ may be
chosen to be $H$ unitary according to the representation $T$ in
Eq.~(\ref{eq03}), i.e.\
\[\VV^\dagger(A)H=H\VV^{-1}(A).\]

Collecting the results obtained, the generators of the interacting Poincar\'e
algebra $p_{1,3}$ have the form
\begin{eqnarray}\label{eq2.11}
\Pi_\mu(A)&=&P_\mu+k_\mu\frac{q}{2k_P}(qA^2-2AP-\slF),\nonumber\\
\sigma_{\mu\nu}(A)&=&s_{\mu\nu}-\frac{q}{k_P}\Bigg(\frac{q}{2k_P}A^2
  (\eta_{\mu\rho}k_\nu-\eta_{\nu\rho}k_\mu)k_\sigma+\strut\nonumber\\&&\strut
  +\eta_{\mu\rho}(k_\nu A_\sigma-k_\sigma A_\nu)
  -\eta_{\nu\rho}(k_\mu A_\sigma-k_\sigma A_\mu)+\strut\nonumber\\&&\strut
  -\frac{q}{k_P}(k_\mu A_\nu-k_\nu A_\mu)k_\rho A_\sigma\Bigg)s^{\rho\sigma},\\
\xi_\mu(A)&=&x_\mu-\frac{q}{2k_P}\Big[x_\mu,\int d\xi(qA^2-2AP)-\slG\Big]
  \nonumber
\end{eqnarray}
where $\slF\equiv F_{\mu\nu}s^{\mu\nu}$ and $\slG\equiv G_{\mu\nu}s^{\mu\nu}$.
The transformed first Casimir operator $\Pi^2(A)$ reads
\begin{equation}\label{eq2.12}
\Pi^2(A)=D^2(A)-q\slF
\end{equation}
where $D_\mu(A)=P_\mu-qA_\mu$. The explicit form of the transformed
Pauli--Lubanski vector $\Omega_\mu(A)$ is
\begin{eqnarray}\label{eq2.13}
\Omega_\mu(A)&=&W_\mu-\frac{q}{2k_P}\epsilon_{\mu\nu\rho\sigma}
  \left\{\eta^{\nu\alpha}\left(\frac{q}{2k_P}A^2k^\rho k^\beta
  +G^{\rho\beta}\right)-\frac{q}{2k_P}G^{\nu\rho}G^{\alpha\beta}\right\}
  s_{\alpha\beta}P^\sigma+\strut\nonumber\\&&\strut
  +\frac{q}{4k_P}\epsilon_{\mu\nu\rho\sigma}k^\sigma\eta^{\nu\alpha}
  \left(\eta^{\rho\beta}-\frac{2q}{k_P}G^{\rho\beta}\right)s_{\alpha\beta}
  \left(qA^2-2AP-\slF\right)
\end{eqnarray}
which yields the transformed second Casimir operator
\begin{eqnarray}\label{eq2.14}
\Omega^2(A)&=&-\frac12s^2D^2
  +s^{\sigma\alpha}s_{\sigma\beta}D_\alpha D^\beta
  +\frac12qs^2\slF+\strut\nonumber\\&&\strut
  -\frac{q}{2k_P}\left\{(k_\alpha s^{\alpha\sigma}\slF)s_{\sigma\beta}
  +s_{\sigma\beta}(k_\alpha s^{\alpha\sigma}\slF)\right\}D^\beta
  +\frac{q^2}{4k_P}(k_\alpha s^{\alpha\sigma}\slF)
  (k^\beta s_{\beta\sigma}\slF)+\strut\nonumber\\&&\strut
  -\frac{iq}{2k_P}(k^\alpha\slF s_{\alpha\beta})D^\beta
  -\frac{iq}{2k_P}(k_\alpha s^{\alpha\sigma})(k^\beta s_{\beta\sigma}\slF').
\end{eqnarray}

\subsection{A nonminimal coupling}
Considering the nonsingular transformation of Dirac-type wave equation
\begin{equation}
\VV(A):(\gamma^\mu P_\mu-m)\psi=0\quad\rightarrow\quad
(\Gamma^\mu(A)\Pi_\mu(A)-m)\Psi(A)=0,
\end{equation}
with the help of Eq.~(\ref{eq2.10}) the dynamical counterparts to the operator
$P_\mu=i\partial_\mu$ can be calculated to be
$\Pi_\mu(A)=\VV(A)P_\mu\VV^{-1}(A)$,
\begin{eqnarray}
P_\mu&\rightarrow&\Pi_\mu(A)=P_\mu+k_\mu\frac{q}{2k_P}(qA^2-2AP-\slF),\\[4pt]
P^2&\rightarrow&\Pi^2(A)=(P-qA)^2-q\slF\label{P2trans}
\end{eqnarray}
($\slF\equiv s^{\mu\nu}F_{\mu\nu}$) while the dynamical counterpart to
$\gamma^\mu$ is given by $\Gamma^\mu(A)=\VV(A)\gamma^\mu\VV^{-1}(A)$,
\begin{equation}
\Gamma^\mu(A)={V^\mu}_\nu(A)\gamma^\nu=\gamma^\mu-\frac{q}{k_P}
  \left(\frac{q}{2k_P}A^2k^\mu k^\nu+G^{\mu\nu}\right)\gamma^\nu.
\end{equation}
In terms of $\Pi_\mu(A)$ and $\Gamma^\mu(A)$ we have
\begin{equation}
\DD^d(A)\Psi(A)=(\Gamma^\mu(A)\Pi_\mu(A)-m)\Psi(A)=0.
\end{equation}
However, expressed in terms of $D_\mu=P_\mu-qA_\mu$ and $\gamma^\mu$, we obtain
\begin{equation}\label{eq3.5}
\DD^d(A)\Psi(A)\equiv\left(\gamma^\mu D_\mu-\frac q{2k_P}\slk\slF-m
  \right)\Psi(A)=0
\end{equation}
where $\slk\equiv\gamma^\mu k_\mu$. This interaction is nonminimal. However,
as we have shown before, it is determined completely by the claim of gauge
invariance.

\vspace{7pt}
Note that due to the antimutation of the $\gamma$-matrices, in the spin-1/2
case the dynamical interaction in Eq.~(\ref{eq3.5}) reduces to the minimal
coupling. However, in order to obtain the correct values of the gyromagnetic
factor, in some cases the (phenomenological) Pauli term
$\gamma_\mu\gamma_\nu F^{\mu\nu}$ has to be added by hand to the minimal
coupling of the Dirac equation (see also Ref.~\cite{Sakurai:1993}, p.~109). In
the case of plane waves the exact solution of this (supplemented) Dirac
equation as given by Chakrabarti~\cite{Chakrabarti:1968zz} obeys the same
gauge invariance condition $\Psi(A+\partial\lambda)=G(\lambda)\Psi(A)$. This
property is found also in the book by Fried~\cite{Fried:1990}.

\vspace{7pt}
Moreover, in the path integral representation of the (nonrelativistic)
Schr\"odinger quantum mechanics the Feynman propagator for any external
electromagnetic field (as an operator $\OO$ on the wave function $\Psi$) is
gauge invariant, i.e.\ the diagram
\begin{center}\begin{tabular}{rlcr}
$\OO\Psi:$&$A$&$\longrightarrow$&$\OO(A)\Psi(A)$\\
$\downarrow\quad$&
$\ \downarrow\rlap{$\lambda$}$&&$\downarrow\rlap{$G(\lambda)$}\qquad$\\
$\OO\Psi^\lambda:$&$A+\partial\lambda$&$\longrightarrow$
  &$G(\lambda)\OO(A)\Psi(A)$\\
\end{tabular}\end{center}
is commutative, $\OO(A+\partial\lambda)G(\lambda)=G(\lambda)\OO(A)$ due to
Eq.~(\ref{kom1}) on the $\psi$-space, and the wave function transforms
as~(\ref{Psitrans}).

\subsection{Local phase transformation}
For the physical quantities $k_\mu$ and $F_{\mu\nu}$ in the model introduced
before in Eqs.~(\ref{eq2.8}) and~(\ref{eq2.9i}) the external (unquantized)
field is acting on the particle without reaction of the particle on the field.
The identification of the elementary particle system with the Poincar\'e group
invariants in Eqs.~(\ref{eq2.12}) and~(\ref{eq2.14}) leads to the equations
\begin{eqnarray}
(P^2-m^2)\psi=0&\rightarrow&(\Pi^2(A)-m^2)\Psi
  =(D^2-qF_{\mu\nu}s^{\mu\nu}-m^2)\Psi=0,\label{eq3.9}\\[7pt]
\left(W^2+m^2s(s+1)\right)\psi=0&\rightarrow&\left(\Omega^2(A)+m^2s(s+1)\right)
  \Psi=0.\label{eq3.10}
\end{eqnarray}
These two equations must be satisfied by any field in the presence of the
plane-wave field. As a consequence of Eq.~(\ref{eq3.9}) the gyromagnetic factor
is $g=2$ and the Bargmann-Michel-Telegdi equation for the four-polarization
vector of the particle takes its simplest form in the proper time frame of the
particle~\cite{Ots:2001xn}.

\vspace{7pt}
Finally, as a consequence of the explicit form~(\ref{eq2.10}), the associated
transformation of the evolution operator $\VV(A)$ under the local gauge
transformation for the plane-wave field,
\begin{equation}\label{eq2.15}
A_\mu(\xi)\rightarrow A_\mu(\xi)+\partial_\mu\lambda(\xi)
\end{equation}
becomes
\begin{equation}\label{eq2.16}
\VV(A)\rightarrow\VV(A+\partial\lambda)
  =e^{-iq\lambda}\VV(A).
\end{equation}
We conclude that the phase transformation is a {\em consequence\/} of the
gauge transformation. This should hold not only for the particular case of
plane waves as analysed explicitly in this section but also for a general
solution $\VV(A)$.

\section{The Rarita--Schwinger equation in the framework of a dynamical
  interaction}
The spin-3/2 field may be described entirely in terms of the vector-bispinor
$\Psi_\mu$ corresponding to the representation of the proper Lorentz group
\begin{equation}\label{eq4.1i}
\left({\textstyle\frac12,\frac12}\right)\otimes\left({\textstyle
  \left(\frac12,0\right)\oplus\left(0,\frac12\right)}\right)
  ={\textstyle\left(1,\frac12\right)\oplus\left(\frac12,1\right)
  \oplus\left(\frac12,0\right)\oplus\left(0,\frac12\right)}.
\end{equation}
The transformation rule according to Eq.~(\ref{eq01}) is
\begin{equation}\label{eq4.1ii}
\left(\tau(\lambda)\psi\right)_\mu(p)=\Lambda_{\mu\nu}T_D(\Lambda)
  \psi^\nu(\Lambda^{-1}p)
\end{equation}
where $T_D(\Lambda)$ is the Dirac representation of the Lorentz group. The
generators of the representation are
\begin{eqnarray}\label{eq4.2}
s_{\mu\nu}&=&-ie_{\mu\nu}\otimes\oone_D+\oone_P\otimes s_{D\mu\nu}
  \ =\nonumber\\
  &=&i\left(-\frac12\eta_{\mu\nu}+E_{\mu\nu}\otimes\oone_D
  -E_{\nu\mu}\otimes\oone_D+\frac12\oone_P\otimes\gamma_\mu\gamma_\nu\right)
\end{eqnarray}
where the indices $P$ and $D$ stand for the Proca and Dirac parts of the
direct product in Eq.~(\ref{eq4.1i}). Here the 16 matrices $E_{\mu\nu}$
generate the Weyl's basis of the set of $4\times 4$ matrices,
\[(E_{\mu\nu})_{\rho\sigma}=\eta_{\mu\rho}\eta_{\nu\sigma},\qquad
E_{\mu\nu}E_{\rho\sigma}=\eta_{\nu\rho}E_{\mu\sigma},\]
and $e_{\mu\nu}=-E_{\mu\nu}+E_{\nu\mu}$ for the Lorentz generators of the
vector representation. The $SO_3$ decomposition of the
representation~(\ref{eq4.1i}) is
\begin{equation}\label{eq4.3}
2D^{(3/2)}\oplus 4D^{(1/2)}.
\end{equation}
Therefore, the representation of the Poincar\'e group contains spins
$3/2$ and $1/2$. The Pauli--Lubanski vector reads
\begin{equation}\label{eq4.4}
W_\mu=i\epsilon_{\mu\rho\sigma\nu}\left(E^{\rho\sigma}\otimes\oone_D
  +\frac14\oone_P\otimes\gamma^\rho\gamma^\sigma\right)P^\nu
\end{equation}
and its square
\begin{equation}\label{eq4.5}
W^2=-\frac{15}4P^2+P^2(E^{\mu\nu}\otimes\gamma_\mu\gamma_\nu)
  +P_\mu P^\nu(E^{\mu\rho}\otimes\gamma_\rho\gamma_\nu
    +E^{\rho\mu}\otimes\gamma_\nu\gamma_\rho).
\end{equation}
Note that
\begin{eqnarray}
(W^2)^2&=&-\frac94P^2\Bigg\{-\frac{15}4P^2
  +P^2(E^{\mu\nu}\otimes\gamma_\mu\gamma_\nu)+\strut\nonumber\\&&\strut
  +P_\mu P^\nu(E^{\mu\rho}\otimes\gamma_\rho\gamma_\nu)
  +P_\mu P^\nu(E^{\rho\mu}\otimes\gamma_\nu\gamma_\rho)+\frac58P^2\Bigg\}
  \ =\nonumber\\
  &=&-2s^2P^2\left(W^2+\frac{s^2-1}2P^2\right)\Bigg|_{s=3/2}
\end{eqnarray}
is a pure spin-3/2 object which enables us to construct the Poincar\'e
covariant mass ($m$) and spin ($j$) projectors
($j=3/2$, $1/2$)~\cite{Napsuciale:2006wr}. The free spin-$3/2$ particle
Rarita--Schwinger equation is given as
\begin{eqnarray}
(P_\nu\gamma^\nu-m)\psi^\mu&=&0,\label{eq4.6i}\\[7pt]
\gamma_\mu\psi^\mu&=&0.\label{eq4.6ii}
\end{eqnarray}
The other constraints
\begin{eqnarray}
(P^2-m^2)\psi^\mu&=&0,\label{eq4.7}\\[7pt]
P_\mu\psi^\mu&=&0\label{eq4.8}
\end{eqnarray}
turn out to be a consequence of Eqs.~(\ref{eq4.6i}) and~(\ref{eq4.6ii}). It is
interesting to note that the static condition~(\ref{eq4.6ii}) and the dynamic
condition~(\ref{eq4.8}) together eliminate the spin-1/2 state completely,
i.e.\ the equations
\[(P^2-m^2)\psi^\mu=0,\qquad\gamma_\mu\psi^\mu=P_\mu\psi^\mu=0\]
with $\psi_\mu$ transforming according to Eq.~(\ref{eq4.1ii}) gives a theory
for spin-3/2 states. Indeed, using the explicit form of $W^2$ in
Eq.~(\ref{eq4.5}) it is easy to see that under the constraints~(\ref{eq4.6ii})
and~(\ref{eq4.8}) we obtain
\begin{equation}\label{eq4.9}
W^2\psi=-\frac{15}4P^2\psi=-s(s+1)P^2\psi\Big|_{s=3/2}.
\end{equation}
Therefore, Eqs.~(\ref{eq4.6i}) and~(\ref{eq4.6ii}) describe indeed a single
particle of mass $m$ and spin $3/2$.

\vspace{7pt}
The dynamical interaction is obtained in the way described in Sec.~3. Taking
into account the explicit form~(\ref{eq4.2}) of the generators $s_{\mu\nu}$,
the transformation $\VV(A)$ in Eq.~(\ref{eq2.10}) becomes
\begin{eqnarray}\label{eq4.10}
\VV_{\rm RS}(A)&=&\exp\left(-\frac{iq}{k_P}\int(AP-\frac q2A^2)\right)
  (\oone_P\otimes\oone_D)\times\strut\nonumber\\&&\strut
  \times\left\{\oone_P-\frac q{k_P}\left(G_{\rho\sigma}-\frac q{2k_P}
  (G^2)_{\rho\sigma}\right)E^{\rho\sigma}\right\}\otimes
  \left\{\oone_D+\frac q{4k_P}G^{\rho\sigma}\gamma_\rho\gamma_\sigma\right\}.
\end{eqnarray}
A straightforward calculation yields
\begin{eqnarray}\label{eq4.11}
P_\mu&\rightarrow&\Pi_\mu(A)=\left(P_\mu+k_\mu\frac q{2k_P}(qA^2-2AP)
  \right)(\oone_P\otimes\oone_D)+\strut\nonumber\\&&\strut
  -k_\mu\frac{iq}{k_P}F_{\rho\sigma}(E^{\rho\sigma}\otimes\oone_D)
  -k_\mu\frac{iq}{4k_P}F_{\rho\sigma}(\oone_P\otimes\gamma^\rho\gamma^\sigma),
  \nonumber\\[7pt]
s_{\mu\nu}&\rightarrow&\sigma_{\mu\nu}(A)=-i\left(\frac12\eta_{\mu\nu}
  +\frac q{k_P}G_{\mu\nu}\right)(\oone_P\otimes\oone_D)
  +\strut\nonumber\\&&\strut
  +i\Bigg\{-\eta_{\mu\rho}\eta_{\nu\sigma}
  +\frac q{k_P}(\eta_{\mu\rho}G_{\nu\sigma}-\eta_{\nu\rho}G_{\mu\sigma})
  +\strut\nonumber\\&&\strut
  -\frac{q^2}{2k_P^2}\left(\eta_{\mu\rho}(G^2)_{\nu\rho}
  -\eta_{\nu\rho}(G^2)_{\mu\sigma}+G_{\mu\nu}G_{\rho\sigma}\right)\Bigg\}
  \left(e^{\rho\sigma}\otimes\oone_D
  -\frac12\oone_P\otimes\gamma^\rho\gamma^\sigma\right),\nonumber\\[7pt]
W_\mu&\rightarrow&\Omega_\mu(A)=-\frac{iq}{2k_P}\epsilon_{\mu\nu\rho\sigma}
  k^\nu A^\rho P^\sigma\oone_P\otimes\oone_D+\strut\nonumber\\&&\strut
  -\frac i2\epsilon_{\mu\nu\rho\sigma}\Bigg\{\left({\eta^\nu}_\alpha
  {\eta^\rho}_\beta
  -\frac{q}{k_P}{\eta^\nu}_\alpha{G^\rho}_\beta
  +\frac{q^2}{2k_P^2}G^{\nu\rho}G_{\alpha\beta}
  -\frac{q^2A^2}{2k_P^2}k^\rho k_\beta{\eta^\nu}_\alpha\right)P^\sigma
  +\strut\nonumber\\&&\strut\qquad
  +\frac{q}{2k_P}(qA^2-2AP)k^\sigma{\eta^\nu}_\alpha\left({\eta^\rho}_\beta
  +\frac{2q}{k_P}A^\rho k_\beta\right)\Bigg\}
  \Big\{e^{\alpha\beta}\otimes\oone_D
  -\frac12\oone_P\otimes\gamma^\alpha\gamma^\beta\Big\}
  +\strut\nonumber\\&&\strut
  +\frac{q}{4k_P}\epsilon_{\mu\nu\rho\sigma}k^\sigma{\eta^\nu}_\alpha
  \left({\eta^\rho}_\beta+\frac{2q}{k_P}A^\rho k_\beta\right)F_{\lambda\tau}
  \times\strut\nonumber\\&&\strut\qquad\times
  \Big\{e^{\alpha\beta}e^{\lambda\tau}\otimes\oone_D
  -\frac12e^{\alpha\beta}\otimes\gamma^\lambda\gamma^\tau
  -\frac12e^{\lambda\tau}\otimes\gamma^\alpha\gamma^\beta
  +\frac14\oone_P\otimes\gamma^\alpha\gamma^\beta\gamma^\lambda\gamma^\tau
  \Big\}.
\end{eqnarray}
The two Casimir invariants of the dynamical Poincar\'e algebra are
\begin{equation}\label{eq4.12}
P^2\ \rightarrow\ \Pi^2(A)=D^2(A)-2iqF^{\rho\sigma}
  \Big\{(E_{\rho\sigma}\otimes\oone_D)
  +\frac14(\oone_P\otimes\gamma_\sigma)\Big\}
\end{equation}
and
\begin{eqnarray}\label{eq4.13}
W^2&\rightarrow&\Omega^2(A)=\left(\frac92
  +(e^{\rho\sigma}\otimes\gamma_{\rho\sigma})\right)D^2
  +\strut\nonumber\\&&\strut
  +\frac12\left(-4(E_{\alpha\beta}\otimes\oone_D)
  +{e_\alpha}^\rho\otimes\gamma_{\rho\beta}
  +{e_\beta}^\rho\otimes\gamma_{\rho\alpha}\right)D^\alpha D^\beta
  +\strut\nonumber\\&&\strut
  -\frac{iq}{2k_P}k^\tau F^{\rho\sigma}\Bigg\{
  -\frac32(h_{\tau\beta}\otimes\gamma_{\rho\sigma})
  +(h_{\tau\rho}\otimes\gamma_{\sigma\beta})+\strut\nonumber\\&&\strut
  -\eta_{\sigma\beta}({h_\rho}^\alpha\otimes\gamma_{\alpha\tau})
  -\frac i2\epsilon_{\rho\sigma\alpha\beta}({e_\tau}^\alpha\times\gamma^5)
  \Bigg\}D^\beta+\strut\nonumber\\&&\strut
  +iqF^{\rho\sigma}\Bigg\{-16(E_{\rho\sigma}\otimes\oone_D)
  -\frac{29}8(\oone_P\otimes\gamma_{\rho\sigma})+\strut\nonumber\\&&\strut
  -6({e^\alpha}_\sigma\otimes\gamma_{\rho\alpha})
  -i\epsilon_{\rho\sigma\alpha\beta}(E^{\alpha\beta}\otimes\gamma_5)\Bigg\}
  +\strut\nonumber\\&&\strut
  -\frac q{k_P}k^\alpha k^\beta F^{\prime\rho\sigma}(E_{\alpha\beta}
  \otimes\gamma_{\rho\sigma})
\end{eqnarray}
where we used the abbreviations $\gamma_{\mu\nu}\equiv\gamma_\mu\gamma_\nu$
and $h_{\mu\nu}\equiv E_{\mu\nu}+E_{\nu\mu}$. Applying the operator
$\VV_{\rm RS}(A)$ to the Rarita--Schwinger equation~(\ref{eq4.6i})
and~(\ref{eq4.6ii}) one obtains
\begin{eqnarray}
\left\{(D^\mu\gamma_\mu-m)\eta_{\rho\sigma}-\frac{iq}{k_P}(k^\mu\gamma_\mu)
  F_{\rho\sigma}\right\}\Psi^\sigma&=&0,\label{eq4.14i}\\[7pt]
\gamma_\mu\Psi^\mu&=&0\label{eq4.14ii}
\end{eqnarray}
where $\Psi(x,A)=\VV_{\rm RS}(x,A)\psi(x)$.

\vspace{7pt}
Equation(\ref{eq4.14i}) is the true equation of motion containing all
derivatives $D_\mu\Psi_\sigma$. The static constraint~(\ref{eq4.14ii})
survives the dynamical interaction and eliminates all superfluous spin-1/2
components. As a consequence the other constraints are the Feynman--Gell-Mann
equation
\begin{equation}\label{eq4.15}
\left\{(\slD^2-m^2)\eta_{\mu\rho}-2iqF_{\mu\rho}\right\}\Psi^\rho=0
\end{equation}
and the kinematical constraint
\begin{equation}\label{eq4.16}
\left\{D_\mu-\frac{iq}{4k_P}(F^{\rho\sigma}\gamma_\rho\gamma_\sigma)k_\mu
  \right\}\Psi^\mu=0.
\end{equation}
Note that as in the free case the ``dynamical'' interaction is algebraically
consistent. Moreover, the second order equation~(\ref{eq4.15}) describes the
causal propagation of waves (assuming the continuity of the first order
derivatives of $\Psi$).

\section{Conclusions and outlook}
Based on the Lorentz--Poincar\'e connection we showed that an external
electromagnetic field $A$ can be introduced most consistently by using the
nonsingular transformation $\VV(A)$. Imposing the two claims that the
transformation (1.) applies not only to the differential operator $\DD$ of the
equation of motion but to the whole Poincar\'e algebra, and (2.) applied to
the generators $s_{\mu\nu}$ of the Poincar\'e algebra yields a Lorentz-type
transformation, the nonsingular transformation $\VV(A)$ is uniquely defined.
For the case of plane waves we showed this explicitly for the Dirac-type
equation and the Rarita--Schwinger equation. The local phase transformation of
the covariant functions $\psi$ appears {\em as a consequence\/} of the local
gauge transformation. This is opposite to the traditional point of view where
phase transformation and gauge transformation are imposed simultaneously.

\vspace{7pt}
An essential point in our approach is that Lorentz and gauge transformation
are placed on the same foundation. Accordingly, the covariant functions in
the presence of an external electromagnetic field $A$ have to depend
explicitly both on the space-time location $x$ and the field $A$,
$\Psi(x,A)$. The field $A$, therefore, has to be understood as coordinate. In
a forthcoming publication we will quantize this system. On the other hand, we
are inspired by the success of the realization of the nonsingular
transformation $\VV(A)$ for the plane-wave case. In a sequel of this paper we
will generalize this to the more general situation of an arbitrary
electromagnetic field $A$.

\subsection*{Acknowledgements}
The work is supported by the Estonian target financed Projects No.~0182647s04
and No.~0180056s09 and by the Estonian Science Foundation under Grants
No.~6216 and No.~8769. S.G. acknowledges the support by the Deutsche
Forschungsgemeinschaft (DFG) under Grant No.~436~EST~17/1/06.

\newpage

\begin{appendix}

\section{Algebraic consistency, locality and causality}
\setcounter{equation}{0}\def\theequation{A\arabic{equation}}
In this Appendix we deal in detail with problems of consistency and causality
related to the introduction of an interacting electromagnetic field into
higher-spin theories, as first mentioned by Velo and
Zwanziger~\cite{Velo:1969bt}.

\subsection{Algebraic inconsistency}
The term ``algebraic inconsistency'' was coined by Velo and
Zwanziger~\cite{Velo:1969bt} and explained explicitly in
1971~\cite{Velo:1971}, where the authors showed that the minimal
substitution for the three equations~(\ref{eq4.6i}), (\ref{eq4.6ii})
and~(\ref{eq4.7}) leads to the unwanted constraint
$e\gamma_\mu F^{\mu\nu}\psi_\nu=0$. As only escape from this ``disaster'' they
proposed the method of Fierz and Pauli~\cite{Fierz:1939ix} where an ansatz for
the interaction is used which had to be adjusted to the physical requirements.
Following this method via the second order Klein--Gordon equation, they ended
up with an additional contribution $O(F)$ to the wave function. The same
procedure but based on the first order equation of motion with a more general
ansatz is used by Porrati and Rahman~\cite{Porrati:2009bs}. A possible
nonminimal action term could be constructed explicitly. The procedure was
extended for the application to massive spin-2 bosonic string states by Argyres
and Nappi~\cite{Argyres:1989cu}, while Porrati {\it et
al.}~\cite{Porrati:2010hm} applied the method to string states with arbitrary
high spin and showed that the BRST operator employed by Argyres and Nappi is
not necessary. However, all these applications of the Fierz--Pauli method
still need a consistency check.

With our method we escape from this necessity because the three
equations~(\ref{eq4.6i}), (\ref{eq4.6ii}) and~(\ref{eq4.7}) are not
independent of each other. Instead, the third is a consequence of the first
two. In applying the nonsingular transformation $\VV_{\rm RS}(A)$ to these two
equations we end up with Eqs.~(\ref{eq4.14i}) and~(\ref{eq4.14ii}).
Equations (\ref{eq4.15}) and~(\ref{eq4.16}) are a consequence and, therefore,
evidently consistent with Eqs.~(\ref{eq4.14i}) and~(\ref{eq4.14ii}). 

\subsection{Locality}
Our results should in principal be comparable with the results of
Ref.~\cite{Porrati:2009bs}. What makes it difficult to perform this
cross-check is the nature of the operator $1/k_P$. As the differential
operator $k_P\equiv k_\mu P^\mu$ commutes with all other operators of the
representation space of the Poincar\'e group, so does $1/k_P$. Therefore,
according to Schur's lemma both operators are diagonal. As reciprocal of a
differential operator, $1/k_P$ need not be local. However, as stressed by
Chakrabarti~\cite{Chakrabarti:1968zz}, Beers and Nickle~\cite{Beers:1972xt},
and later by Brown and Kowalski~\cite{Brown:1983bc,Brown:1984hy}, as applied
to eigen states to the Poincar\'e group, the operator still turns out to
be local and contributes to the local Lorentz transformation $V_{\mu\nu}(A)$.
For the near future we hope to overcome these difficulties and perform the
comparison with the results given by Porrati and Rahman~\cite{Porrati:2009bs}
as well as with Deser {\it et al.}~\cite{Deser:2000dz}.

\subsection{Causality}
Because the Poincar\'e group takes care of the space-time structure of the
result and, therefore, the causality, there is no need to show the causality
of the result explicitly. In this context it is worth stressing that the
Velo--Zwanziger problem is not the final word. As explained by
Cox~\cite{Cox:1989zz}, the constraint analysis of Velo and
Zwanziger~\cite{Velo:1969bt} is not complete because the ``true equation of
motion'' still does not determine the time derivatives. In completing the
analysis, instead of acausality Cox finds a loss of degrees of freedom.

Using our method, an explicit check for causality was performed in
Ref.~\cite{Saar:1999ez} by analyzing the characteristic surfaces
(see e.g.~\cite{Courant:1962}), as it was employed starting from Velo and
Zwanziger (for a detailed explanation see Ref.~\cite{Zwanziger:1977ud}) up to
recent works of Porrati {\em et al.\/}~\cite{Porrati:2010hm}. Our result for
the normal vector $n_\mu$ obeying~\cite{Saar:1999ez}
\begin{equation}
\Delta(n)=\pfrac12^4\left(n^2\right)^8
\end{equation}
shows that every characteristic surface is a lightcone and the propagation,
therefore, is causal.

\end{appendix}

\end{document}